\documentclass[final,5p,times,twocolumn,sort&compress]{elsarticle}

\usepackage{graphicx}
\usepackage{bm}
\usepackage{amsmath}
\usepackage{amssymb}

\newcommand{\etal}{\emph{et al.}~}
\newcommand{\ie}{\emph{i.e.}~}
\newcommand{\eg}{\emph{e.g.}~}
\newcommand{\Eq}{Eq.~}
\newcommand{\Eqs}{Eqs.~}
\newcommand{\Fig}{Fig.~}
\newcommand{\Figs}{Figs.~}

\journal{Ultramicroscopy}

\begin{document}

\begin{frontmatter}

\title{Factors limiting quantitative phase retrieval in atomic-resolution differential phase contrast scanning transmission electron microscopy using  a segmented detector}

\author[a]{T. Mawson}
\author[a]{D.J. Taplin}
\author[b]{H.G. Brown}
\author[c]{L. Clark}
\author[d,e]{R. Ishikawa}
\author[d]{T. Seki}
\author[d]{Y. Ikuhara}
\author[d]{N. Shibata}
\author[a]{D.M. Paganin}
\author[a]{M.J. Morgan}
\author[f,g]{M. Weyland}
\author[a,f]{T.C. Petersen}
\author[a]{S.D. Findlay}

\address[a]{School of Physics and Astronomy, Monash University, Clayton, Victoria 3800, Australia}
\address[b]{Ian Holmes Imaging Center, Bio21 Molecular Science and Biotechnology Institute, University of Melbourne, Victoria 3010, Australia}
\address[c]{School of Chemical and Process Engineering, University of Leeds, Leeds LS2 9JT, UK}
\address[d]{Institute of Engineering Innovation, University of Tokyo, Tokyo 113-8656, Japan}
\address[e]{PRESTO, Japan Science and Technology Agency, Kawaguchi, Saitama 3320012, Japan}
\address[f]{Monash Centre for Electron Microscopy, Monash University, Clayton, Victoria 3800, Australia}
\address[g]{Department of Materials Science and Engineering, Monash University, Clayton, Victoria 3800, Australia}


\begin{abstract}

Quantitative differential phase contrast imaging of materials in atomic-resolution scanning transmission electron microscopy using segmented detectors is limited by various factors, including coherent and incoherent aberrations, detector positioning and uniformity, and scan-distortion. By comparing experimental case studies of monolayer and few-layer graphene with image simulations, we explore which parameters require the most precise characterisation for reliable and quantitative interpretation of the reconstructed phases. Coherent and incoherent lens aberrations are found to have the most significant impact. For images over a large field of view, the impact of noise and non-periodic boundary conditions are appreciable, but in this case study have less of an impact than artefacts introduced by beam deflections coupling to beam scanning (imperfect tilt-shift purity).

\end{abstract}

%\begin{keyword}
%\end{keyword}

\end{frontmatter}

\section{Introduction}

The principal direct-imaging technique in high-resolution scanning transmission electron microscopy (STEM) has long been incoherent, Z-contrast, high-angle annular dark field (HAADF) imaging. However, in recent years coherent phase contrast techniques have enjoyed a resurgence, particularly differential phase contrast (DPC). The underlying principle has long been known \cite{DdL1,R28}: the specimen electromagnetic field induces directional intensity redistribution between suitably arranged detectors in the diffraction plane which gives rise to image contrast. This has been used to good effect in probing long range electromagnetic fields \cite{CMM1,UZ1,LSJWWSZ1,BHLRBZ1}. Atomic resolution DPC STEM has since been achieved using both segmented detectors \cite{SFKSKI1,close2015towards,lazic2016phase,yucelen2018phase} and fast-readout pixel detectors \cite{Muller_2014_KBSGLVZSR,hachtel2018sub,muller2018atomic,gao2019real}, demonstrating that it has good sensitivity to both light and heavy elements, and that it is dose-efficient.

Pixel detectors allow for a high-accuracy calculation of the first moment, or centre-of-mass, of the diffraction pattern intensity. For a true phase object this first moment image is equal to the exact gradient field of the phase convolved with the probe intensity \cite{WC1,Lubk_2015_Z,lazic2016phase,clark2018probing}. As such, most work seeking high-precision, quantitative DPC imaging has used pixel detectors \cite{muller2017measurement,gao2019real}.\footnote{For completeness, it should be noted that more elaborate ptychographic methods offer further strategies for overcoming many of the limiting factors considered here \cite{jiang2018electron,chen2021electron}, though they require a pixel detector and a greater degree of post-processing.} For segmented detectors, the issue of field sensitivity has been much explored at lower resolution \cite{zweck2016detector,schwarzhuber2017achievable,wu2017correlative,lee2019influence,haas2019direct}. At atomic resolution, explorations of factors limiting the accuracy of reconstructions have tended to focus on the fact that segmented detectors only provide an approximation to the first moment \cite{close2015towards,lazic2016phase,muller2019comparison} and can introduce asymmetric artefacts \cite{lazic2016phase,Yang_2014_PN}. However, it has also been shown that the difference between the exact and approximate first moments may be smaller than the impact of other factors (especially dynamical diffraction) on the interpretability of DPC STEM images \cite{close2015towards}, and that if the sample is weakly scattering then the asymmetry in the phase contrast transfer function can largely be corrected \cite{seki2017quantitative}. Similarly, while segmented detector DPC STEM has a weaker phase contrast transfer function than that of a pixel detector \cite{Yang_2014_PN}, at least for some segmented detector configurations the noise-normalised phase contrast transfer function is only marginally weaker \cite{seki2018theoretical}. Segmented detectors also readily allow for live monitoring of DPC signals, \ie displaying them during acquisition at the same scanning speed as that used for HAADF images, facilitating the identification of features of interest and the fine-tuning of the imaging conditions \cite{shibata2017electric,shibata2017direct}.

In this manuscript we explore the relative impact of various instrumental and experimental factors---including probe-forming aperture size, coherent and incoherent aberrations, camera length, detector rotation, dose and scan-distortion---on the quantitative reliability of DPC STEM imaging with a segmented detector. The goal is a better understanding of which instrumental factors require the most precise characterisation. The exploration is framed around an experimental test case using a single STEM instrument and a graphene sample (thereby avoiding complications from multiple scattering). While the detailed findings may thus be somewhat specific to the instrument and system used, we emphasise principles that are expected to be more general. The findings may also apply to pixel detectors, though that geometry facilitates the characterisation of several of the quantities considered.

The paper is structured as follows. In section \ref{sec:instchar} we discuss the instrument characterisation, including estimating incoherent effective source size from the HAADF image. In section \ref{sec:theory} we briefly review the theory for phase reconstruction in DPC-STEM for a weakly scattering object, which informs post-processing choices and the interpretation of imaging artefacts. To explore the quantitative accuracy achieved, in section \ref{sec:repunitavg} we directly compare repeat-unit-averaged data from monolayer graphene against simulation, together with further parameter space exploration in simulation to explore the sensitivity to parameters including camera length and detector misalignment. In section \ref{sec:nonp} we consider larger fields of view, where repeat-unit averaging is not possible and boundary conditions are no longer periodic.

\section{Instrument characterisation}
\label{sec:instchar}

\begin{figure*}[hbt!]
\begin{center}
  \includegraphics*[width=1.8\columnwidth]{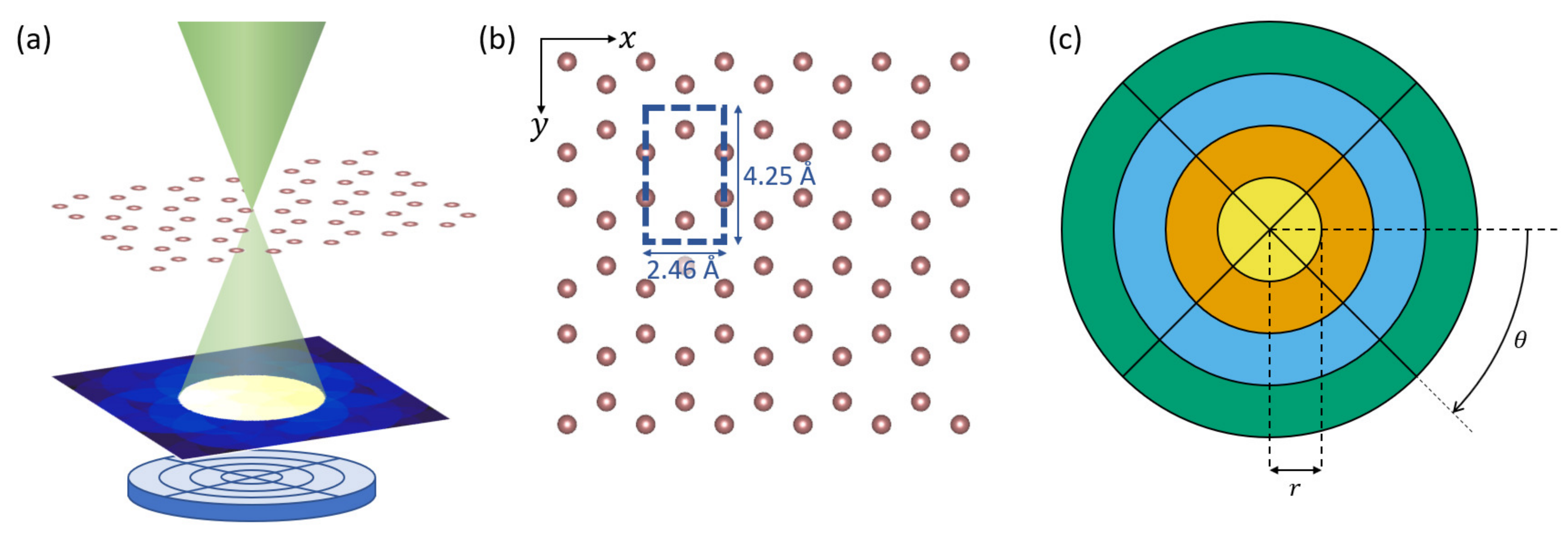}
\caption{(a) DPC STEM schematic, showing an electron probe passing through a graphene monolayer and forming a diffraction pattern, with intensity slightly redistributed by the scattering, falling upon a segmented detector. (b) Schematic of the structure of monolayer graphene, indicating our definition of coordinate axes for later reference. (c) Schematic of the 16-segment detector used. Common colouring has been used to indicate the segments within each ring on the detector. In what follows, we make reference to the rotation of the detector, here denoted by $\theta$, and to the size of the detector (a consequence of the camera length used) as characterised by the radius of the innermost ring, here denoted by $r$. The detector orientation shown here relative to the graphene structure in (b) closely reflects that used in the experimental data shown subsequently. \label{fig:a1}}
\end{center}
\end{figure*}

Segmented detector STEM images were obtained on the JEOL ARM300CF installed at the University of Tokyo, with the microscope operated at 80 kV and a probe-forming aperture semi-angle of 27 mrad. Figure \ref{fig:a1}(a) depicts the idealised set-up, with a STEM probe illuminating a graphene sample, the interaction with which leads to some intensity redistribution in the diffraction plane intensity that falls upon a segmented detector. Figure \ref{fig:a1}(b) shows the structure of an ideal graphene monolayer and defines our coordinate system. As depicted in \Fig \ref{fig:a1}(c), the segmented detector consists of four concentric rings divided into quadrants, such that there are 16 detector segments in all. The widths of each annulus are assumed to be the same (for the innermost ring, effectively a disk, this is the radius $r$ in \Fig \ref{fig:a1}(c)). See Ref.\ \cite{SKFSKI1} for a fuller description of the hardware, and Ishikawa \etal \cite{ishikawa2018direct} for a previous application of this instrument to DPC-STEM of graphene. The camera length was nominally set such that the outer edge of the bright field disk extended out to about the centre of detector ring 3, but we refined that estimate through trial-and-error scaling of the effective detector size ($r$ in \Fig \ref{fig:a1}(c)) as follows.

\begin{figure}[htb!]
\begin{center}
  \includegraphics*[width=0.85\columnwidth]{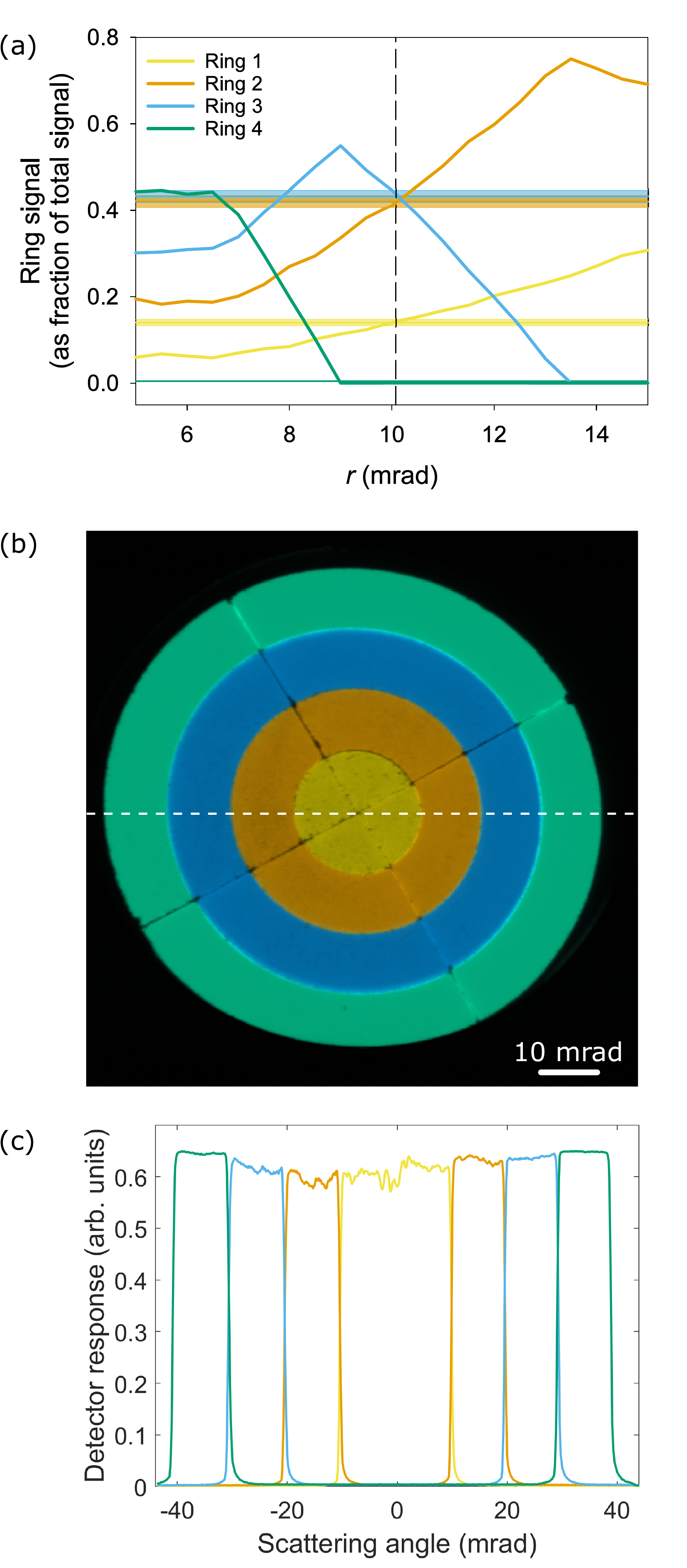}
\caption{(a) Determination of detector span (or equivalently camera length), parameterised using $r$ as defined in \Fig \ref{fig:a1}(c). The thick, solid lines show the simulated ring signals assuming ideal detector response functions and a uniform bright field disk resulting from a probe-forming aperture semiangle of 27 mrad. The thin horizontal bands indicate the $\pm 1\sigma$ uncertainty ranges of the experimentally measured signals in the different rings. Consistent agreement is found for $r\approx 10$ mrad (indicated by  the vertical dashed line). (b) Measured detector response map, coloured to make the connection with \Fig \ref{fig:a1}(c). Scale bar based on the calibration from (a). (c) Line scan across the measured detector response map, taken from the white dashed line in (b). The detector uniformity is good but not perfect, with some ``tails'' in the response function extending beyond the nominal geometric edges of each segment. \label{fig:a2}}
\end{center}
\end{figure}

Figure \ref{fig:a2}(a) plots the average ring signals of simulated segmented detector STEM images as a function of the outer radius of ring 1 ($r$ in \Fig \ref{fig:a1}(c)) and as a fraction of the total detector signal. The calculations assume a monolayer of graphene,\footnote{For specimens where the thickness or structure is less well known and so cannot readily be included in simulation, it would be preferable to use signals recorded without the sample present.} the accelerating voltage and probe-forming aperture semi-angle of the experiment, and ideal detector response. Equivalent experimental data were obtained, and are shown as the thin horizontal bands in \Fig \ref{fig:a2}(a), which for each ring indicate the $\pm 1\sigma$ uncertainty range of the experimentally measured signal (accounting for both the variance in the graphene images and in the black-level for the detectors, the latter obtained from an acquisition of the same duration but with the beam blanked). The intercepts between the simulated lines and experimental values for a ring 1 outer radius of $r=10$ mrad show good consistency for rings 1 to 3, which puts the outer edge of the 27 mrad bright field disk slightly further out than the centre of ring 3. Though for the signal in ring 4 the values in both experiment and simulation are very small for $r=10$ mrad, the experimental signal exceeds that of the simulated signal. This discrepancy is attributed to the non-ideal detector response. Figure \ref{fig:a2}(b) shows the detector response maps, coloured to underscore the connection with \Fig \ref{fig:a1}(c). Figure \ref{fig:a2}(c) shows line scans taken across the white dashed line in \Fig \ref{fig:a2}(b). Some non-uniformity in the detector response is evident as fluctuations in the line scans, but the more significant effect is the ``tails'' in the detector response that extend beyond the nominal geometric edges of each segment.\footnote{The challenge of tightly focusing the probe in the detector plane for the detector response scan may mean \Figs \ref{fig:a2}(b) and (c) overstate this effect, but does not fully explain the effect.} This is most significant for ring 4. Geometrically one would expect no signal in ring 4 in the absence of a sample, and only minimal contribution due to scattering in the presence of monolayer graphene. However, the tails lead to some contribution from electrons within the bright field region. While this explains the lack of quantitative match for ring 4 in \Fig \ref{fig:a2}(a), the effect is nevertheless small. We show later through simulation that neither uncertainty in camera length nor detector non-uniformity are limiting factors in the quantitative accuracy of our measurements.

One factor appreciably impacting quantitative STEM imaging is source incoherence. Temporal incoherence due to chromatic aberration can be incorporated into simulation based on the measured chromatic aberration coefficient, $C_{\rm c} = 0.89$ mm, and the full-width-half-maximum of the spread (assumed Gaussian) of energies emitted from the cold field emission gun tip, $\Delta E = 0.45$ eV. Spatial incoherence is harder to reliably measure, since it depends not purely on instrument characteristics but also on stability factors that may vary from experiment to experiment \cite{Dwyer2012}. Since we are interested in how quantitative segmented detector DPC STEM reconstructions are, to estimate the effective source size describing spatial incoherence we fit not to the bright field images but rather to the HAADF image, following the method of Yamashita \etal \cite{yamashita2015quantitative}.

\begin{figure*}[htb!]
\begin{center}
  \includegraphics*[width=2.0\columnwidth]{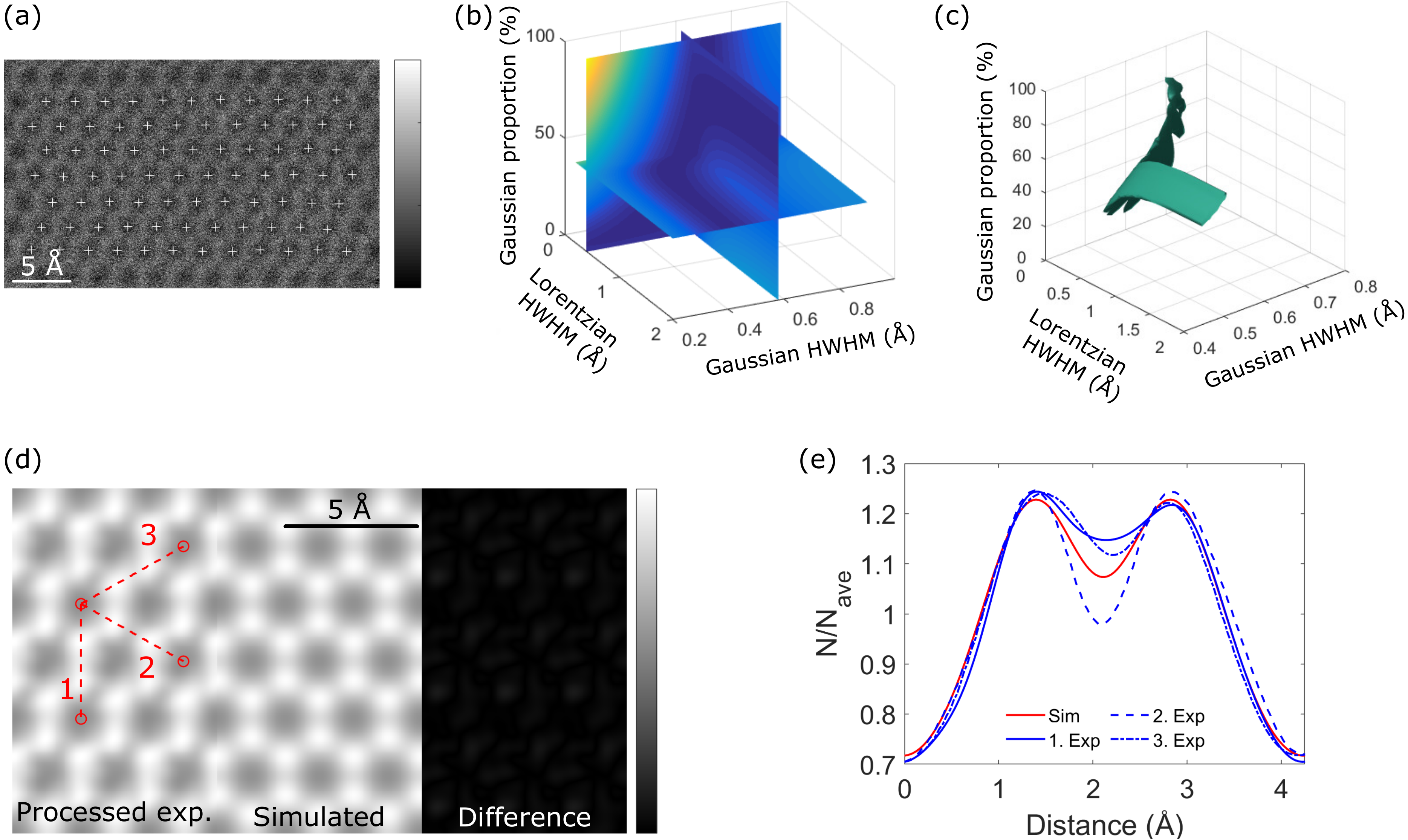}
\caption{(a) Raw HAADF image of monolayer graphene. The array of bright crosses superposed on the estimated centre of the graphene hexagons are used as reference points for repeat-unit averaging. (b) Three orthogonal slices through a Euclidean norm measure of the difference between the repeat-unit-averaged experiment and simulations assuming an incoherent effective source distribution that is a superposition of a Gaussian component (characterised by its half-width-half-maximum) and a Lorentzian component (characterised by its half-width-half-maximum) with variable weighting (characterized by the proportion comprised by the Gaussian). (c) Isosurface of the Euclidean norm measure in close vicinity to the minimum value attained. (d) Repeat-unit-averaged experimental HAADF image (``processed exp.''), the corresponding simulation (assuming a Gaussian half-width-half-maximum of 0.58 {\AA}, a Lorentzian half-width-half-maximum of 0.40 {\AA} and a Gaussian proportion of 42\%) scaled such that the mean value equals that of the experiment, and their difference. All three panels are on the same scale, and the difference is seen to be small. (e) Projected line profiles over the three rectangles indicated in (d), compared against the same profile for the simulation. Despite the profiles being over regions that by symmetry should be structurally equivalent, some variability is seen between the experimental profiles. Nevertheless, they are broadly in agreement with the simulation.
 \label{fig:a3}}
\end{center}
\end{figure*}

Figure \ref{fig:a3}(a) shows a raw HAADF image from a region of monolayer graphene, with crosses marking the centres of the hexagonal rings as determined by inverting the contrast and cross-correlating the result with a Gaussian function of comparable width to that across the hexagon centres. For each triangular region with adjacent such reference points as vertices, local affine transformations are defined that map the recorded pixels to their location within the conventional rectangular (\ie non-primitive) repeating unit cell for graphene, indicated by the blue dashed rectangle in \Fig \ref{fig:a1}(b). Repeat-unit averaging is effected by determination of the Fourier coefficients of the STEM image that best match these values in a least-squares sense, a strategy described in greater detail in Ref. \cite{chen2016practical}. The result is seen on the left panel (``processed exp.'') in \Fig \ref{fig:a3}(d).

A simulated HAADF image incorporating the diffraction limit, chromatic aberration and 47-171 mrad HAADF detector range produces an image in which the carbon atoms are better resolved than they are in the processed experimental image. The difference is attributed to spatial incoherence \cite{Dwyer2012}. Following Yamashita \etal \cite{yamashita2015quantitative}, we assume the incoherent effective source distribution can be described as a linear combination of Gaussian and Lorentzian (more precisely, bivariate Cauchy distribution) components. Figure \ref{fig:a3}(b) shows three orthogonal projections through an $\ell^2$ (Euclidean) norm of the difference between the repeat-unit-averaged experimental data and simulated data as a function of the three free parameters in the effective source distribution. For the present data there is no unequivocal minimum in this parameter sweep. Instead, all points in the rather complex isosurface plotted in \Fig \ref{fig:a3}(c) give essentially the same near-minimum value. Choosing somewhat arbitrarily amongst these values a Gaussian half-width-half-maximum of 0.58 {\AA}, a Lorentzian half-width-half-maximum of 0.40 {\AA} and a Gaussian proportion of 42\% gives the simulated HAADF image in the central panel in \Fig \ref{fig:a3}(d). Visually, the processed experimental and simulated images appear in good agreement, and their difference, shown on the same intensity scale in the right panel in \Fig \ref{fig:a3}(d), is much smaller than either.

A clearer quantitative perspective is given in \Fig \ref{fig:a3}(e) which plots line-scans (averaged over a width of about 0.3 {\AA}) from the lines shown in the left panel in \Fig \ref{fig:a3}(d). Based on the symmetry of the ideal graphene structure, these three line profiles should be identical. This is true for the simulation, but not for the experimental data, since the repeat-unit averaging was based on a (non-primitive) rectangular cell and so the three- and six-fold symmetries are not enforced. That the profiles vary in which peaks and troughs they over-estimate or under-estimate helps clarify why the range of parameter combinations in \Fig \ref{fig:a3}(c) can give similarly good fits: with differences in symmetry such that no fit can be perfect, parameter variations that improve the fitting in one region of the image while worsening it in another region do not appreciably change the overall fit.

The variability among the experimental profiles across structurally-equivalent regions is likely evidence of some asymmetry to the probe not accounted for in the simulations, which is plausible since astigmatism is hard to eliminate completely when the raw data are very noisy. It is also possible that some residual distortion remains in the unit-cell-averaged data, or that the incoherent effective source distribution best describing the experiment is not rotationally symmetric. However, without an independent means of verification we think introducing too many instrumental free parameters into the modelling would be misleading for an exploration of which factors limit attempts at quantitative comparison between experiment and simulation. Therefore, we will persist with the present values while bearing in mind this limitation in instrument characterisation.

\section{Theory of differential phase contrast imaging for weak phase objects}
\label{sec:theory}

Our analysis largely follows that of Seki \etal \cite{seki2017quantitative,seki2021toward}. In what follows, we use CoM (centre-of-mass) to refer to the exact first moment (obtainable to an excellent approximation if one uses a pixel detector) and DPC to refer to approximations to the first moment obtained using a segmented detector.

We write the coherent, diffraction-limited STEM probe wavefield incident upon the sample with probe at position ${\bf R}$ as
\begin{equation}
\psi_{in}({\bf r}_\perp-{\bf R},\Delta f) = \int T_{\Delta f}({\bf k}_\perp) e^{2\pi i {\bf k}_\perp \cdot ({\bf r}_\perp-{\bf R})} d{\bf k}_\perp \;,
 \label{eq:01}
\end{equation}
where ${\bf r}_\perp$ is a coordinate vector in the plane of the specimen, with corresponding Fourier space coordinate vector ${\bf k}_\perp$, and $T_{\Delta f}({\bf k}_\perp)=A({\bf k}_\perp) \exp[-i \chi({\bf k}_\perp,\Delta f)]$ is the lens transfer function, comprising both the aperture function $A({\bf k}_\perp)$ and the aberration function $\chi({\bf k}_\perp,\Delta f)$. Though for simplicity we only explicitly denote the lens defocus value $\Delta f$, other aberrations can readily be included \cite{kirkland2020advanced}.

The effect of spatial and temporal incoherence on any ideal, coherent STEM image $I_{\rm coh}({\bf R},\Delta f)$ is given by \cite{nellist1994beyond}
\begin{equation}
I({\bf R}) = I_{\rm coh}({\bf R},\Delta f) \otimes_{\bf R} S({\bf R}) \otimes_{\Delta f} \sigma({\Delta f}) \;,
 \label{eq:02}
\end{equation}
where $S({\bf R})$ is the effective source size characterising spatial incoherence, $\sigma(\Delta f)$ is the effective defocus spread distribution resulting from temporal incoherence, and $\otimes$ denotes convolution, which in cases of potential ambiguity is subscripted with the coordinate over which convolution is performed. As per the left side of Eq. (\ref{eq:02}), for notational simplicity we tend to drop the explicit $\Delta f$ dependence of the recorded STEM image.

We write the intensity at diffraction plane coordinate ${\bf k}_\perp$ when the STEM probe is at position ${\bf R}$ as $I({\bf k}_\perp,{\bf R})$. For a given detector response function $D({\bf k}_\perp)$, the resultant STEM image is given by
\begin{equation}
I({\bf R}) = \int I({\bf k}_\perp,{\bf R}) D({\bf k}_\perp) d{\bf k}_\perp \;.
 \label{eq:03}
\end{equation}

An ideal first-moment or ``centre-of-mass'' detector would have detector response\footnote{Real detectors have finite  extent, but M{\"u}ller-Caspary \etal \cite{muller2017measurement} have shown that this tends not to adversely affect the evaluation of the first moment in atomic-resolution DPC STEM.}
\begin{equation}
D_{{\rm CoM},\alpha}({\bf k}_\perp) = k_\alpha \;,
 \label{eq:04}
\end{equation}
where $\alpha$ denotes $x$ or $y$ such that $k_\alpha$ is the $\alpha$ component of ${\bf k}_\perp$. In the (strong) phase object approximation, the STEM image resulting from such a detector response can be related to the phase $\phi({\bf r}_\perp)$ of the sample transmission function (which being proportional to the projected specimen potential is the mathematical embodiment of the sample structure) via\footnote{Since our analysis \emph{assumes} a rotationally-symmetric probe (despite it probably not being true of the experiment), we have opted to simplify the presentation by writing \Eq (\ref{eq:05}) and those that follow from it as convolutions. However, as per Lazi{\'c} \etal \cite{lazic2016phase}, to correctly model for a non-rotationally symmetric probe \Eq (\ref{eq:05}) must rather be written
\begin{equation*}
I_{{\rm CoM},\alpha}({\bf R}) = \frac{1}{2\pi}|\psi_{in}({\bf R},\Delta f)|^2\star_{\bf R} \left[\partial_\alpha \phi({\bf R})\right] \otimes_{\bf R} S({\bf R}) \otimes_{\Delta f} \sigma({\Delta f})
\end{equation*}
where $\star$ denotes cross-correlation.}
\begin{align}
I_{{\rm CoM},\alpha}({\bf R}) &= \frac{1}{2\pi}\left[\partial_\alpha \phi({\bf R})\right] \otimes_{\bf R} |\psi_{in}({\bf R},\Delta f)|^2 \otimes_{\bf R} S({\bf R}) \otimes_{\Delta f} \sigma({\Delta f}) \nonumber \\ &= \frac{1}{2\pi}\partial_\alpha \left[\phi({\bf R}) \otimes P_{in}({\bf R}) \otimes S({\bf R}) \right] \;, 
 \label{eq:05}
\end{align}
where $\partial_\alpha$ denotes partial differentiation in the $\alpha$ direction, the second line follows because differentiation and convolution are effected by the commuting multiplication of functions in Fourier space, and for convenience we have defined
\begin{equation}
P_{in}({\bf R}) \stackrel{\rm def}{=} |\psi_{in}({\bf R},\Delta f)|^2 \otimes_{\Delta f} \sigma({\Delta f}) \;.
 \label{eq:06}
\end{equation}
Using the Fourier derivative theorem, it can be shown that
\begin{align}
\phi({\bf R}) &\otimes P_{in}({\bf R}) \otimes S({\bf R}) \qquad\qquad\qquad\qquad\qquad \nonumber \\ &= {\mathcal F}^{-1} \left\{ \frac{ {\mathcal F}\left[I_{{\rm CoM},x}({\bf R})\right] + i {\mathcal F}\left[I_{{\rm CoM},y}({\bf R})\right] }{ i(K_x+iK_y) } \right\} \;,
 \label{eq:07}
\end{align}
where ${\mathcal F}$ denotes Fourier transform from real-space coordinate ${\bf R}$ to Fourier space coordinate ${\bf K}$ (with Cartesian components $K_x$ and $K_y$) \cite{arnison2004linear,kottler2007two,de2008quantitative}. This expression is essentially equivalent to the inverse Laplacian applied to $\partial_x I_x + \partial_y I_y$ assuming a consistency of mixed partial derivatives.\footnote{To appreciate the equivalence, note that the meaning of division by a complex quantity is such that
\begin{equation*}
\frac{ \tilde{I}_{x} + i \tilde{I}_y }{K_x+iK_y} \frac{K_x-iK_y}{K_x-iK_y} = \frac{ K_x \tilde{I}_{x} + K_y \tilde{I}_y  }{ K^2_x+K^2_y} + i \frac{ K_x \tilde{I}_{y} - K_y \tilde{I}_x  }{ K^2_x+K^2_y} 
\end{equation*}
(using the shorthand $\tilde{I}_{x} = {\mathcal F}\left[I_{{\rm CoM},x}({\bf R})\right]$ and likewise for the $y$ component). The first term, which after the inverse Fourier transform in \Eq (\ref{eq:07}) is purely real, is precisely the Fourier transform solution of Poisson's equation \cite{lazic2016phase,ishizuka2017boundary}. The second term would be zero if the measured data satisfied the integrability condition $\partial_y I_x = \partial_x I_y$, but, if not, then after the inverse Fourier transform in \Eq (\ref{eq:07}) it gives a purely imaginary contribution. Taking just the real part of \Eq (\ref{eq:07}) is the projection onto the solution that best satisfies integrability in a least squares sense \cite{frankot1988method}. As we do invariably take the real part, this approach is essentially equivalent to that of Refs.\ \cite{lazic2016phase,ishizuka2017boundary}, though being able to inspect the imaginary component to assess what might loosely be considered the deviation from integrability in the data (due, for instance, to noise and scan distortion) can be a useful check on data quality.} In principle one might deconvolve the instrument contributions to isolate the sample phase $\phi({\bf R})$, but, as discussed later, stability of deconvolution is often an issue.

Using a segmented detector one can determine an approximate first moment via the effective detector response
\begin{equation}
D_{{\rm DPC},\alpha}({\bf k}_\perp) = \left\{ \begin{array}{cl} k^{\rm CoM}_{\alpha,j} & \;\; {\rm if} \; {\bf k}_\perp \; {\rm lies \; within \; the \;} j^{\rm th} \; {\rm segment} \\ 0 & \;\; {\rm otherwise} \end{array} \right.
 \label{eq:08}
\end{equation}
where $k^{\rm CoM}_{\alpha,j}$ denotes the $\alpha$ component of the geometric centre-of-mass of detector segment $j$. In the \emph{weak} phase object approximation, the STEM image resulting from this detector response may be written\footnote{Unlike \Eq (\ref{eq:05}), \Eqs (\ref{eq:09}) and (\ref{eq:10}) are correct even if the probe is not rotationally symmetric.}
\begin{equation}
I_{{\rm DPC},\alpha}({\bf R}) = \frac{1}{2\pi}\left[\partial_\alpha \phi({\bf R})\right] \otimes h_\alpha({\bf R}) \otimes S({\bf R}) 
 \label{eq:09}
\end{equation}
where
\begin{align}
h_\alpha({\bf K}) &= {\mathcal F} \left[ h_\alpha({\bf R}) \right] \nonumber \\ &= \frac{1}{K_\alpha} \Bigg\{ \int \Big[ T^*_{\Delta f}({\bf k}_\perp) T_{\Delta f}({\bf k}_\perp-{\bf K}) \nonumber \\ & \qquad\qquad - T_{\Delta f}({\bf k}_\perp) T^*_{\Delta f}({\bf k}_\perp+{\bf K})\Big] D_{{\rm DPC},\alpha}({\bf k}_\perp) d{\bf k}_\perp \Bigg\} \nonumber \\ & \qquad\qquad\qquad \otimes_{\Delta f} \sigma({\Delta f}) \;.
 \label{eq:10}
\end{align}

The similarity in functional form between \Eq (\ref{eq:09}) and the first line of \Eq (\ref{eq:05}) is evident. Strictly, \Eq (\ref{eq:09}) only holds in the weak object approximation, though Seki \etal \cite{seki2017quantitative} show that for some detector configurations the transfer function is broadly similar to that of the ideal first moment detector, and Close \etal \cite{close2015towards} show that even for thicker samples the segmented detector can give a fairly good approximation to the ideal first moment.

As it stands, \Eq (\ref{eq:09}) is not quite amenable to the inversion step between \Eqs (\ref{eq:05}) and (\ref{eq:07}) due to the $\alpha$ dependence of $h_\alpha({\bf R},\Delta f)$. One might seek to deconvolve that contribution, but as already mentioned deconvolution raises stability concerns. However, since
\begin{align}
I_{{\rm DPC},x}({\bf R}) &\otimes h_y({\bf R}) \nonumber \\  
&= \frac{1}{2\pi}\partial_x \left[ \phi({\bf R}) \otimes h_x({\bf R}) \otimes h_y({\bf R}) \otimes S({\bf R})\right] \;, \nonumber \\ 
I_{{\rm DPC},y}({\bf R}) &\otimes h_x({\bf R}) \nonumber \\ 
&= \frac{1}{2\pi}\partial_y \left[ \phi({\bf R}) \otimes h_x({\bf R}) \otimes h_y({\bf R}) \otimes S({\bf R})\right] \;,
 \label{eq:11}
\end{align}
then, in broad analogy with \Eq (\ref{eq:07}), we may write
\begin{align}
\phi({\bf R}) &\otimes h_x({\bf R}) \otimes h_y({\bf R}) \otimes S({\bf R}) \nonumber \\ &= {\mathcal F}^{-1} \Bigg[\frac{1}{i(K_x+iK_y)} \Bigg\{ {\mathcal F}\left[I_{{\rm DPC},x}({\bf R})\otimes h_y({\bf R})\right] \nonumber \\  & \qquad\qquad + i {\mathcal F}\left[I_{{\rm DPC},y}({\bf R})\otimes h_x({\bf R})\right] \Bigg\} \Bigg] \;. 
 \label{eq:12}
\end{align}

\begin{figure}[htb!]
\begin{center}
  \includegraphics*[width=1.0\columnwidth]{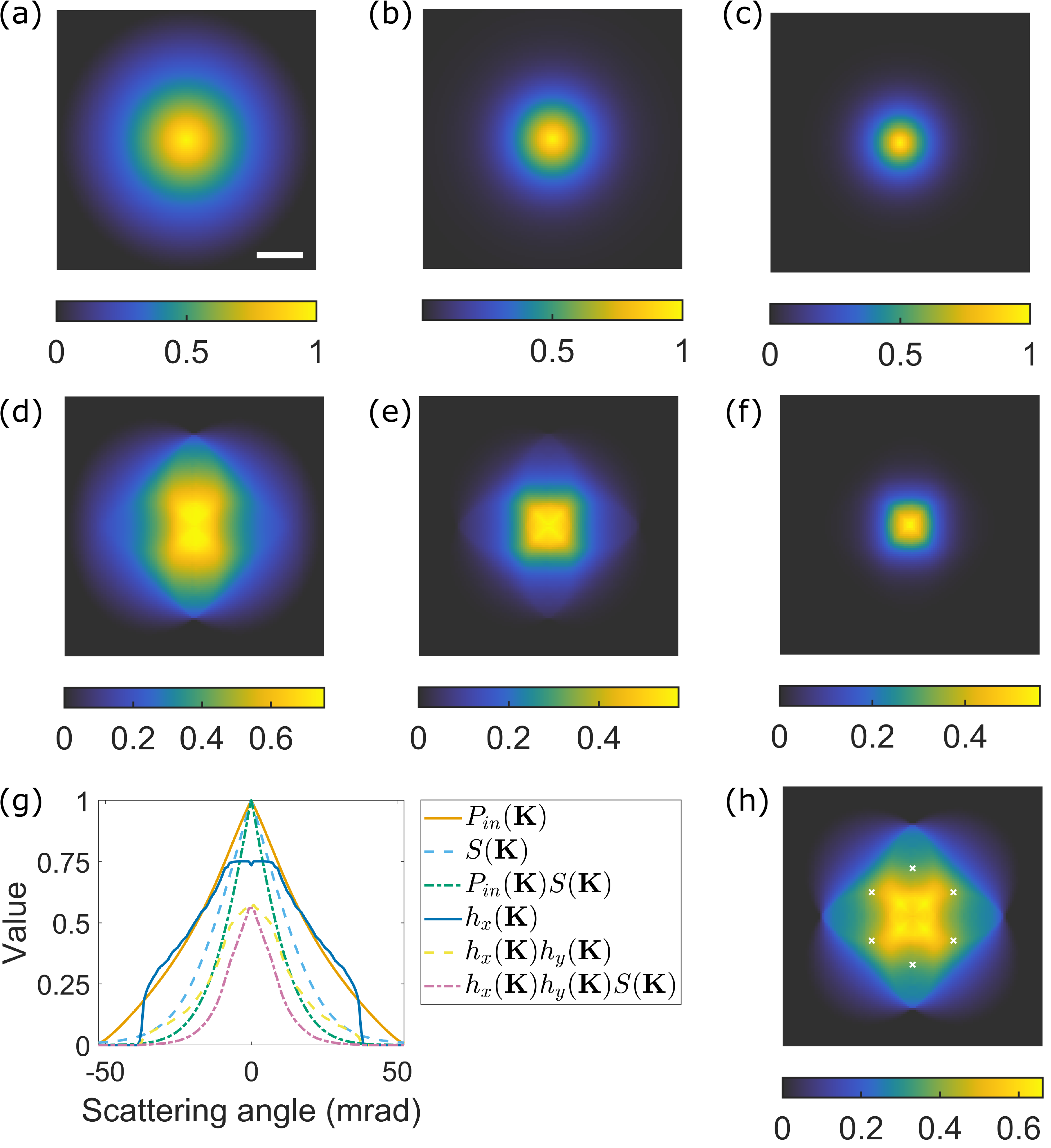}
\caption{Fourier space transfer functions: (a) $P_{in}({\bf K})$, (b) $S({\bf K})$, (c) $P_{in}({\bf K}) S({\bf K})$, (d) $h_x({\bf K})$, (e), $h_x({\bf K}) h_y({\bf K})$, and (f) $h_x({\bf K}) h_y({\bf K}) S({\bf K})$. Each image is shown on its own colour scale to maximise visibility of structure. (g) Line profiles taken vertically through the centre of the transfer functions in (a)-(f). (h) Ratio $h_x({\bf K}) h_y({\bf K}) / P_{in}({\bf K})$, set equal to zero outside the bandwidth limit. The locations of the innermost Bragg spots are indicated by white crosses. The scale bar in (a) is 20 mrad and common to all images.
 \label{fig:a4}}
\end{center}
\end{figure}

Figure \ref{fig:a4} compares some of these transfer functions. Figure \ref{fig:a4}(a) shows $P_{\rm in}({\bf K})$, the Fourier transform of Eq. (\ref{eq:06}). Figure \ref{fig:a4}(b) shows $S({\bf K})$, the Fourier transform of $S({\bf R})$. Figure \ref{fig:a4}(c) shows their product $P_{\rm in}({\bf K})S({\bf K})$, the transfer function for ideal first-moment imaging (see \Eq (\ref{eq:05})). Figure \ref{fig:a4}(d) shows $h_x({\bf K})$. To the extent that \Eq (\ref{eq:09}) is the segmented detector equivalent of \Eq (\ref{eq:05}), the comparison between \Fig \ref{fig:a4}(d) and \Fig \ref{fig:a4}(a) shows that the phase contrast transfer function for the segmented detector has notable asymmetry. However, since we use the pseudo-symmetrisation strategy of Eq. (\ref{eq:11}), the more relevant comparison is with $h_x({\bf K}) h_y({\bf K})$, shown in \Fig \ref{fig:a4}(e): an improvement, but some asymmetry remains. Note that the range of $h_x({\bf K}) h_y({\bf K})$ is about half that of $P_{\rm in}({\bf K})$ (as can be better seen in the line profiles of Fig. \ref{fig:a4}(g)): the contrast transfer function for the segmented detector is weaker than that of an ideal first-moment detector \cite{Yang_2014_PN}. When spatial incoherence in included, the segmented detector case in \Fig \ref{fig:a4}(f) and the ideal first-moment detector case in \Fig \ref{fig:a4}(c) appear qualitatively quite similar, but this is more reflective of the extent to which the large effective source here suppresses the contrast transfer at high frequencies. A better comparison is the ratio of phase contrast transfer functions $h_x({\bf K}) h_y({\bf K}) / P_{in}({\bf K})$ shown in \Fig \ref{fig:a4}(h) (set equal to zero outside bandwidth limit), where the asymmetry remains notable. However, because the phase contrast transfer function in \Fig \ref{fig:a4}(f) falls off rapidly at higher frequencies (due mainly to the spatial incoherence), only the lower spatial frequencies contribute appreciably. The locations of these spatial frequencies relative to the ratio in \Fig \ref{fig:a4}(h) are shown by the white crosses. Despite the asymmetry of the contrast transfer function overall, for the present configuration the transfer has very similar values at those points. As such, for this particular sample and source partial coherence, the implications of using a segmented detector rather than an ideal first-moment detector for quantitative DPC STEM will primarily be a slightly reduced dose efficiency.

\section{Quantitative reliability of differential phase contrast applied to repeat-unit-averaged data}
\label{sec:repunitavg}

Since monolayer graphene is a weak scatterer we expect both the phase object approximation underpinning \Eq (\ref{eq:07}) and the weak phase object approximation underpinning \Eq (\ref{eq:12}) to apply. It is therefore expected that \Eq (\ref{eq:12}), which includes the detailed detector geometry through the transfer functions $ h_\alpha({\bf R})$, will give a more accurate description than \Eq (\ref{eq:07}), which is based on exact first moments that segmented detector measurements only approximate. Nevertheless, in this section we compare both as applied to the segmented detector images obtained from the same monolayer graphene field of view as that of the HAADF image used in \Fig \ref{fig:a3} for effective source size characterisation. In particular, we use the same repeat-unit-averaging process, both to improve the signal-to-noise ratio (which is low because graphene is such a weak scatterer) and so that the effective source size characterisation of section \ref{sec:instchar} applies (since it likely includes a small additional blurring introduced by that averaging). We will explore the quantitative agreement between experiment and simulation (using an independent atom model and the parameterisation due to Waasmaier and Kirfel \cite{waasmaier1995new}) with a view to better understanding which factors may be limiting the agreement and with what care some of the key factors need to be determined.

\begin{figure*}[htb!]
\begin{center}
  \includegraphics*[width=2.0\columnwidth]{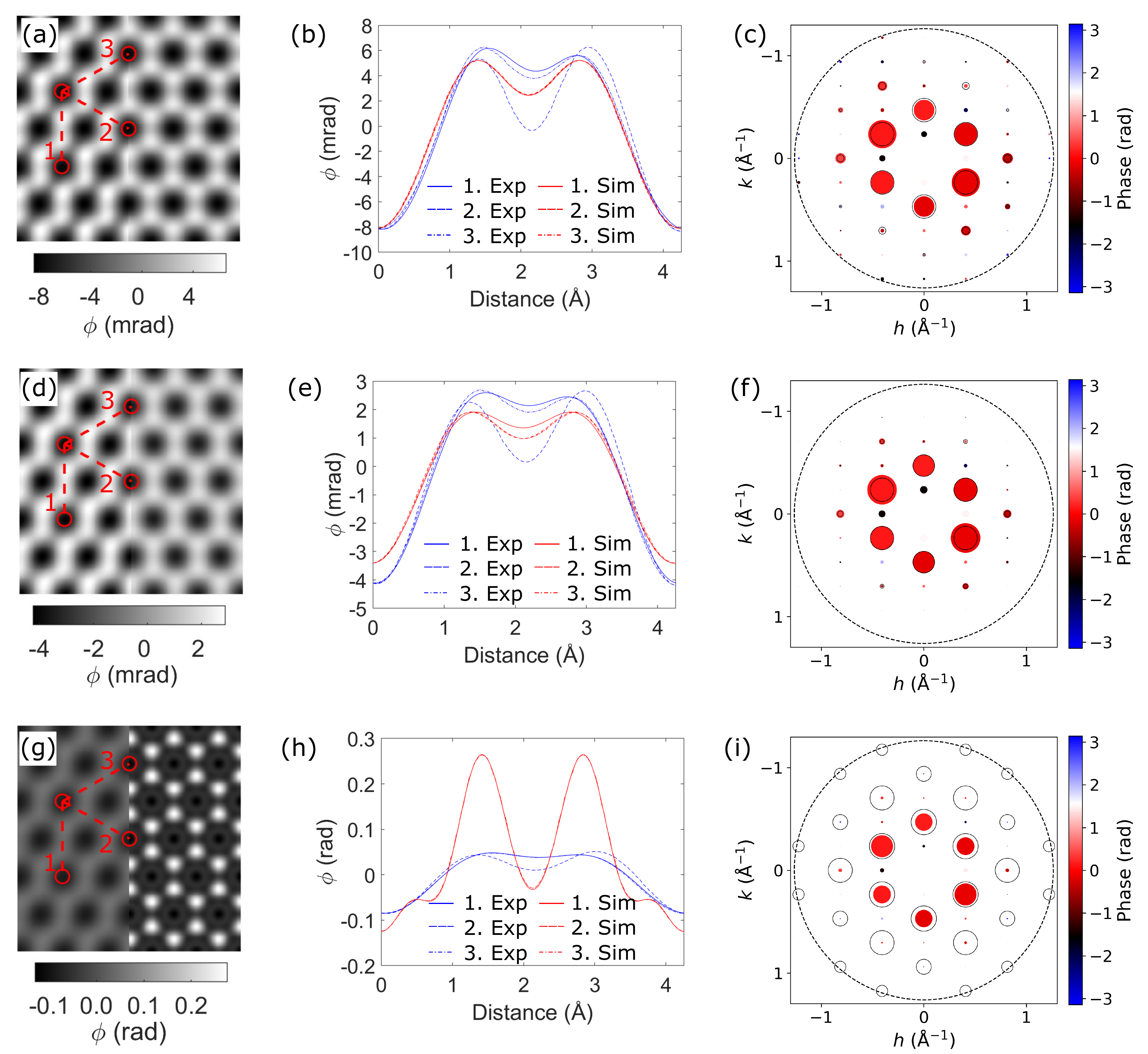}
\caption{Top row: Comparison between the Fourier reconstruction applied to the experimental data following Eq. (\ref{eq:07}) and an independent-atom-model simulation of the expected phase convolved with the total probe intensity distribution, \ie including both coherent and incoherent contributions. The comparison is made as (a) an image, (b) line scans (locations indicated in (a)), and (c) a Fourier coefficient comparison plot. In the Fourier coefficient comparison plot, the area of each circle or disk is proportional to the magnitude of the Fourier coefficient at that point, the disk colour represents the phase difference between the experimentally reconstructed and simulated Fourier coefficients (as given by the colour bar), and the large black dashed-line circle indicates the bandwidth limit, being twice the probe-forming aperture. Middle row: Comparison between the weak object approximation reconstruction of the experimental data following Eq. (\ref{eq:12}) and the simulated phase convolved with the transfer functions (as per the left side of Eq. (\ref{eq:12})). The comparison is made as (d) an image, (e) line scans (locations indicated in (d)), and (f) a Fourier coefficient comparison plot. Bottom row: (g)-(i) Analogous to (d)-(f) but with the transfer functions deconvolved.
 \label{fig:a5}}
\end{center}
\end{figure*}

Figures \ref{fig:a5}(a)-(c) compare the result of the ideal first moment reconstruction of the right side of \Eq (\ref{eq:07}) applied to approximate first moment images (obtained from the data from the segmented detector) against an independent atom model simulation of the expected phase convolved with the source distributions in accordance with the left side of \Eq (\ref{eq:07}). Figure \ref{fig:a5}(a) compares images of this phase side by side on a common scale. Figure \ref{fig:a5}(b) compares line scans (averaged over a width of about 0.3 {\AA}) along the three different lines indicated on \Fig \ref{fig:a5}(a). The symmetry of the structure is such that these line profiles should be identical, as they are for the simulated data (``Sim''), but there are appreciable differences in the experimental profiles. This is very similar to what was seen for the HAADF profiles in \Fig \ref{fig:a3}(e), and is attributed to the same underlying cause: an uncharacterised asymmetry in the coherent and incoherent lens aberrations that is not included in the simulations.

Figure \ref{fig:a5}(c) is a representation of the comparison between Fourier coefficients of the reconstruction as per the right side of \Eq (\ref{eq:07}) and an independent-atom-model simulation of the expected phase convolved with the source distributions in accordance with the left side of \Eq (\ref{eq:07}). The black solid-lined circles have area proportional to the magnitude of the Fourier coefficient at the reciprocal space coordinate at the centre of each circle. (The black dashed-line circle has twice the radius of the bright field disk and indicates the bandwidth limit imposed by the probe-forming aperture.) The coloured disks represent the Fourier coefficients of the phase reconstructed from experimental data, with area proportional to the magnitude of the Fourier coefficient, and colour indicating the phase difference between the experimentally reconstructed and simulated reference Fourier coefficients. Perfect agreement between experiment and simulation would manifest as red disks exactly filling each black solid-lined circle. Broadly consistent with the level of agreement seen in the real-space comparisons in \Figs \ref{fig:a5}(a) and (b), both experiment and simulation in \Fig \ref{fig:a5}(c) show that the six main low-order Fourier coefficients have similar magnitude, and that the small discrepancies in these coefficients are slightly different in the different directions. It also shows small ``super-lattice'' spots in the experimental reconstruction. These arise because the repeat-unit averaging was based on a (non-primitive) rectangular cell, and as such the aforementioned three- and six-fold symmetries expected of ideal graphene are not enforced. The presence of these super-lattice spots is thus evidence of residual asymmetry (which could be due to shot-noise, scan distortion or lens aberration asymmetries, or more likely a combination of all three). However, bearing in mind that the magnitudes of Fourier coefficients are proportional to the disk areas, not radii, these spots represent a small contribution to the total signal.

Figures \ref{fig:a5}(d)-(f) present the same kinds of comparisons, but now between the result of applying the approximate first moment reconstruction of the right side of \Eq (\ref{eq:12}) against an independent atom model simulation of the expected phase convolved with the source distributions in accordance with the left side of \Eq (\ref{eq:12}). The extra convolutions involved in \Eq (\ref{eq:12}) are evident in the reduced phase range in \Figs \ref{fig:a5}(d) and (e), and in the more rapid drop off in circle radius in \Fig \ref{fig:a5}(f). Since we expect the weak phase object approximation to hold well for graphene, this analysis should be more accurate than the ideal first moment analysis applied to segmented detector data. This is evident in the modest differences between some of the different simulation line profiles in \Fig \ref{fig:a5}(e), in qualitative agreement with the experimental data and resulting from the directionality of the segmented detectors meaning that $h_x({\bf R}) \otimes h_y({\bf R})$ is not rotationally symmetric (cf.\ $P_{in}({\bf R})$, which is rotationally symmetric here). It is not so clearly evident in the quantitative agreement: there is some suggestion in \Fig \ref{fig:a5}(f) that the Fourier space comparison agrees slightly better for the low order Fourier coefficients than in \Fig \ref{fig:a5}(c), consistent with the weak phase object formulation more correctly incorporating the geometry of the segmented detector, but conversely the range of the line scans in \Fig \ref{fig:a5}(e) shows proportionally larger discrepancies than in \Fig \ref{fig:a5}(b). Since the weak phase object approximation analysis is the more rigorous approach for this sample, it must be concluded that any better agreement obtained by the ideal first moment analysis is fortuitous, the consequence of compensating errors. All that said, the difference is marginal, suggesting that the approximate first moment is a sufficiently good approximation to the exact first moment that the latter formulation gives a comparably good result here --- it may not suffice if high precision measurements of bonding information in the Fourier coefficients were sought, but it seems quite sufficient for forming visually-interpretable images. We stress again the significant spatial incoherence present here. Had the degree of coherence been higher, such that higher order Fourier coefficients contributed more, the sensitivity to the difference between approximate and exact first moments may have been greater.

Because \Figs \ref{fig:a5}(c) and (f) are in reciprocal space, the convolution theorem implies that the blurring functions manifest as products (in their reciprocal space form). This leads to a strong suppression of the higher order Fourier coefficients. It also means the sample phase in \Figs \ref{fig:a5}(a)-(f) is not a pure specimen property but rather still has imaging system properties folded in. Figures \ref{fig:a5}(g)-(i) attempt to undo this via deconvolution (via the Richardson-Lucy deconvolution algorithm implemented in MATLAB\textsuperscript{\textregistered} ver.\ R2018a). However, it is seen that the results are poor, at least for the present data set. In particular, \Fig \ref{fig:a5}(i) shows that while the deconvolution has worked fairly well on the large, low-order Fourier coefficients, it has been ineffective at retrieving the higher-order Fourier coefficients. Basically, once such terms become comparable with the noise level in the experimental data they can no longer be retrieved with any confidence. Consequently, in what follows all comparisons of experiment against simulation are made without seeking to deconvolve effective source terms to isolate the specimen phase. Note that this is only a strictly valid strategy for quantitative comparison when the effective source terms can be adequately characterised.

\begin{figure*}[htb!]
\begin{center}
  \includegraphics*[width=2.0\columnwidth]{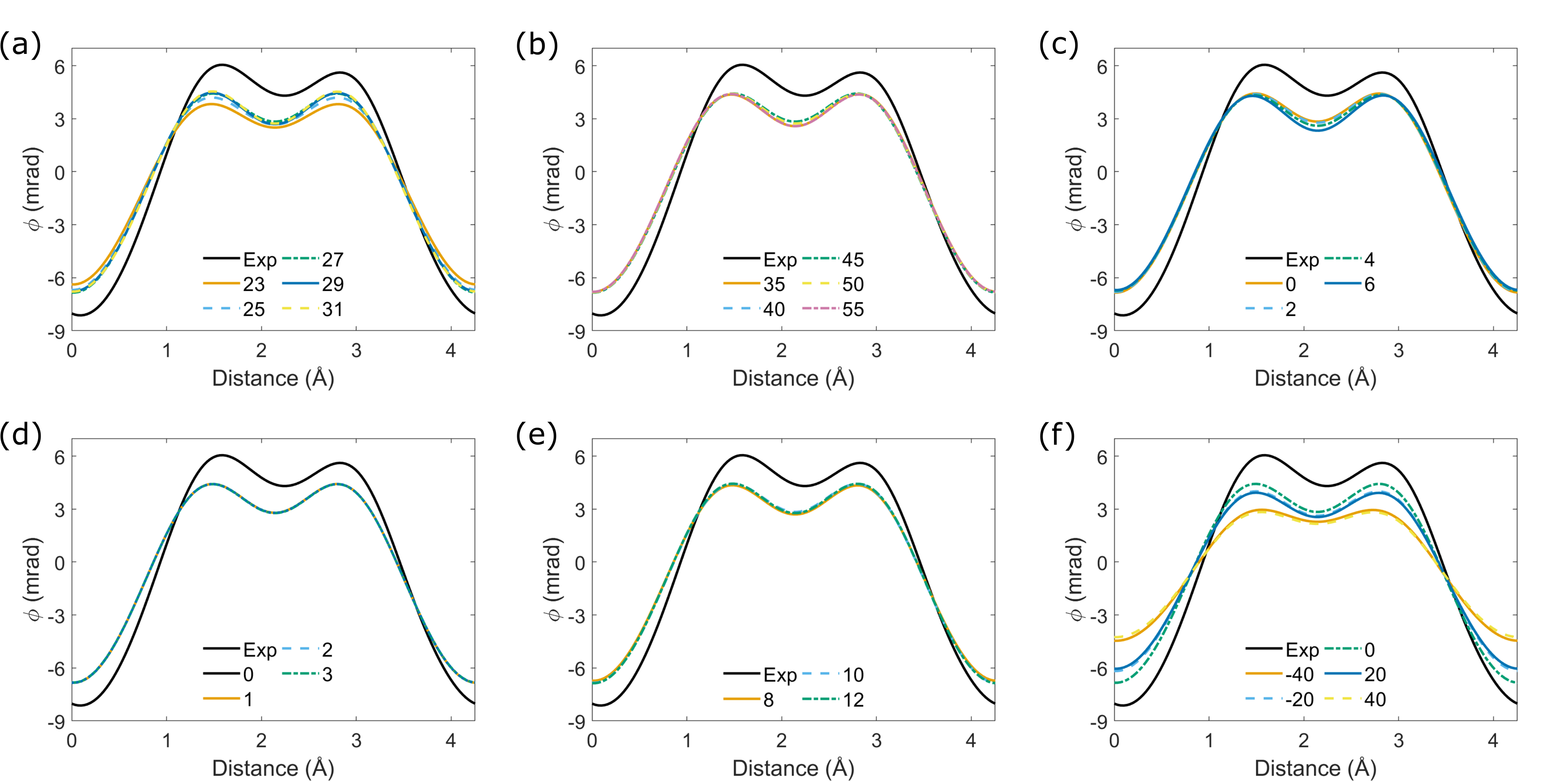}
\caption{Comparison of line profile 1 in \Fig \ref{fig:a5} of the exact first moment reconstruction formulation of Eq. (\ref{eq:07}) applied to the experimental data against those by applying the same reconstruction formulation to simulating segmented detector STEM images for monolayer graphene, varying different experimental parameters one at a time. (a) Varying convergence angle (in mrad). (b) Varying detector rotation angle (in degrees). (c) Varying detector centring (expressed as a shift in mrad along the $y$ direction). (d) Varying the detector uniformity (0--ideal detector response; 1--detector tails extending about 1.5 mrad; 2--detector tails extending about 3 mrad; 3--detector tails extending about 1.5 mrad and non-uniform response at approximately a $\pm 5$\% level). (e) Varying camera length (expressed as a change in $r$ from \Fig \ref{fig:a1}(c), the detector radius of the innermost detector ring, in mrad). (f) Varying defocus (in Angstrom units).
 \label{fig:a6}}
\end{center}
\end{figure*}

We have attributed the deviations from the expected 6-fold symmetry of graphene in \Figs \ref{fig:a5}(a)-(f) to uncharacterised asymmetry in the source distributions (both coherent and incoherent contributions). In seeking to make highly quantitative comparisons, it is useful to understand the effect various instrumental parameters have on these reconstructions, and therefore how much effort and care need be spent on characterising them. 
The line profiles in \Fig \ref{fig:a6} explore the impact of six key parameters: convergence angle, detector rotation, detector centring, detector non-uniformity, camera length (parameterised here by detector radius), and defocus. These profiles are obtained by simulating segmented detector STEM images for monolayer graphene and reconstructing the phase using the ideal first moment analysis formulation of \Eq (\ref{eq:07}). Note that these parameters can also affect pixel detector data, though that geometry may more readily facilitate characterisation and correction.

Figure \ref{fig:a6}(a) explores convergence angle (given in milliradian in the legend). The line scan corresponds to that labelled 1 in \Fig \ref{fig:a5}(a), and the experimental profile from that region (denoted ``Exp'') is also shown. The nominal experimental convergence angle is 27 mrad. Figure \ref{fig:a6}(a) shows that convergence angles in the range 25-31 mrad would produce essentially the same results.\footnote{In practice, changing the aperture size may change the probe current, which in turn would change the noise level, but infinite dose was assumed to simulate \Fig \ref{fig:a6}. In section \ref{sec:nonp} we explore some consequences of a realistic noise level.} Even the 23 mrad case, more perceptibly different than the rest, is far less different from the 27 mrad simulation than the experimental profile is from any of them. Since convergence angle is fairly easy to characterise within a milliradian or so, we conclude than it is unnecessary to characterise it any more precisely.

Figure \ref{fig:a6}(b) explores detector rotation (given in degrees). The best estimate for the experiment is 49.9$^\circ$, but the simulations in \Figs \ref{fig:a5}(d)-(i) assumed 45$^\circ$. Figure \ref{fig:a6}(b) shows this discrepancy does not produce any perceptible change, and indeed nor would further errors in rotation of at least 10$^\circ$. This too, then, is a parameter that can readily be characterised to a much higher precision than is needed for quantitative work.

Figure \ref{fig:a6}(c) explores detector miscentring (given in mrad displacement off centre along the $y$ direction). Displacements of a few milliradian, though a significant fraction of the 10 mrad polar width of each ring, are seen to cause only a very small change. We will see in section \ref{sec:nonp} that over a large field of view an offset in the detector centre can have a larger effect, though one that can be largely suppressed in post-processing. But for small repeat unit and with periodic boundary conditions enforced, the impact is very small.

Figure \ref{fig:a6}(d) explores detector non-uniformity. Case 0 corresponds to an idealised detector response as per \Fig \ref{fig:a1}(c). Cases 1 and 2 correspond to detector responses with tails, reminiscent of what is seen in \Fig \ref{fig:a2}(c), obtained by convolving the ideal detector responses with a Gaussian of standard deviation 0.5 {\AA} and 1.0 {\AA}, respectively. Case 3 also has tails as per convolution with a Gaussian of standard deviation 0.5 {\AA}, but in addition has a non-uniform response reminiscent of that seen in \Fig \ref{fig:a2}(c). The latter effect was obtained by multiplying the detector response map by a low-pass filtered image with Gaussian noise of mean 1 and standard deviation 0.08, which produced fluctuations at around the $\pm 5$\% level. In \Fig \ref{fig:a6}(d), these plots are all indistinguishable, suggesting that modest deviations from uniformity are also not an appreciable limitation.

Figure \ref{fig:a6}(e) explores camera length as parameterised by the outer radius of ring 1 of the detector (given in milliradian). As per \Fig \ref{fig:a2}, the value for the present experiment is close to 10 mrad, and that figure shows the precision in that value to be less than a milliradian. Figure \ref{fig:a6}(e) shows that virtually indistinguishable results would have been obtained had we assumed the radius of the innermost ring on the detector was either 8 mrad or 12 mrad. This is a little surprising, since the former would change which ring of the detector the outer edge of the bright field disk falls within, but underscores that segmented detectors can indeed make reliable estimates of the centre of mass in many cases.

Figure \ref{fig:a6}(f) explores probe defocus (given in angstrom). For reference, the (full-width-half-maximum) depth-of-focus of a perfectly coherent 80 keV probe with 27 mrad convergence angle would be about 100 {\AA}, and the full-width-half-maximum of the spread in defocus due to chromatic aberration (included in these simulations) is 50 {\AA}. Figure \ref{fig:a6}(d) shows that an error of 40 {\AA} in the assumed defocus relative to the experiment could produce changes comparable in magnitude (though in this case in the wrong direction) to that of the difference between our ideal reference and the experimental profile. This is consistent with our interpretation that the main source of asymmetry and discrepancies is the inadequate characterisation of the probe, both in terms of coherent lens aberrations and perhaps also in spatial incoherence. We therefore conclude that improving the characterisation of these quantities should be the priority for improving quantification of experimental DPC reconstructions. Other authors have drawn similar conclusions \cite{cao2018theory}.

\section{Further reliability considerations in DPC over large fields of view}
\label{sec:nonp}

\begin{figure*}[htb!]
\begin{center}
  \includegraphics*[width=2.0\columnwidth]{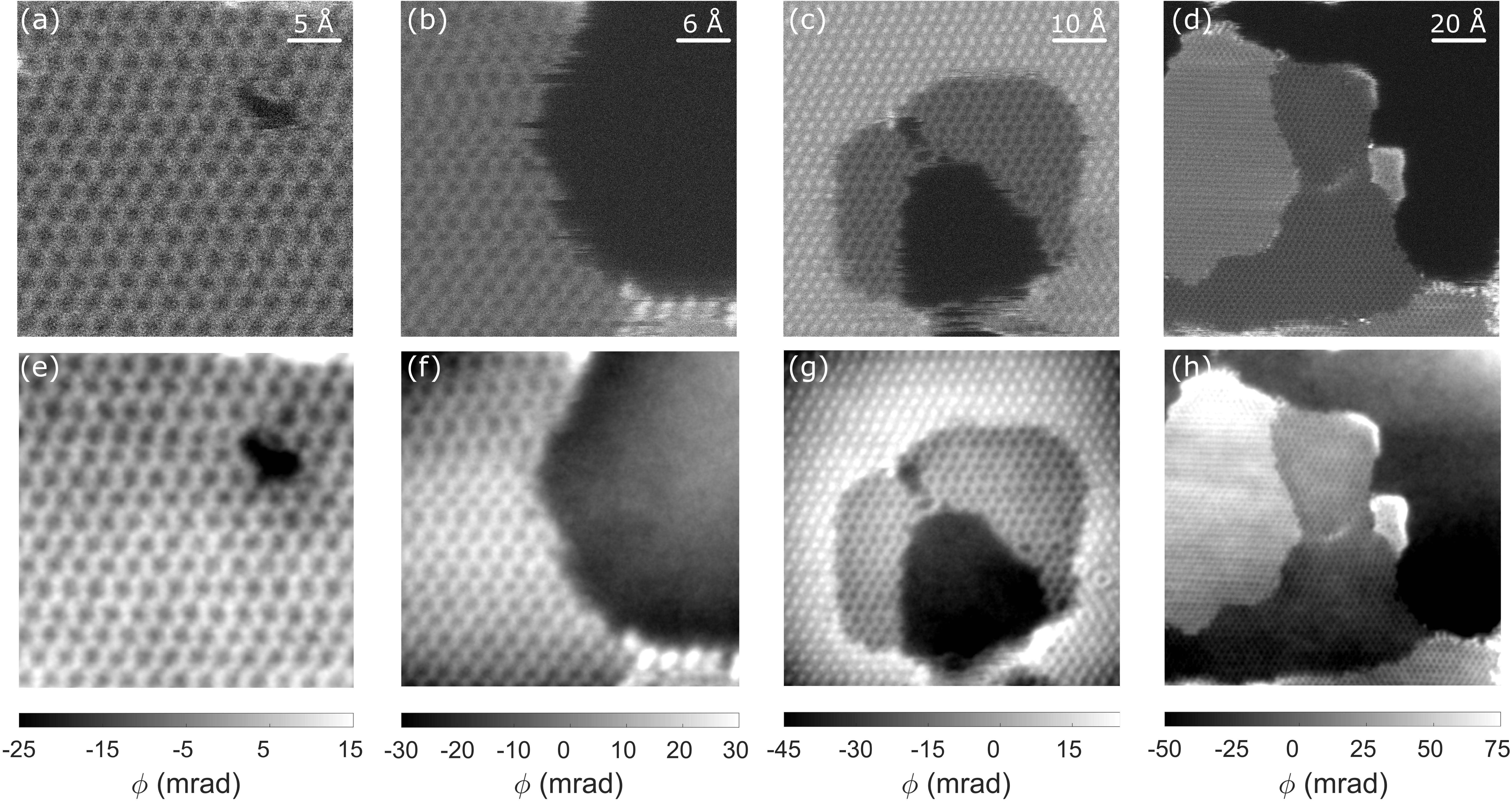}
\caption{(a)-(d) HAADF STEM images of four different regions on the graphene sample. For the same regions (e)-(h) show the reconstructed phases (source blurring effects included) based on the ideal first moment analysis of Eq. (\ref{eq:07}) modified to handle the non-periodic boundary conditions following additional post-processing as summarised in \Fig \ref{fig:a8}. The display range of the reconstructed phase has been truncated slightly to improve the visibility of structure in the phase mid-range.
 \label{fig:a7}}
\end{center}
\end{figure*}

By using repeat-unit-averaged data in the previous section, that exploration largely omits some further factors which might impact DPC reconstruction: noise,\footnote{We explore the impact of the noise level of the present experiment on this form of reconstruction. More systematic explorations of noise / dose issues in DPC STEM can be found in Refs.\ \cite{lazic2016phase,seki2018theoretical,muller2019comparison}.} scan distortion and lack of periodic boundary conditions. The present section will explore the effects of these factors on experimental data from four different regions of a graphene sample at various magnifications, the HAADF images for which are shown in \Figs \ref{fig:a7}(a)-(d). Figures \ref{fig:a7}(a) and (b) contain predominantly monolayer regions within their respective fields of view, whereas \Figs \ref{fig:a7}(c) and (d) show some monolayer and some bilayer regions (bright contrast at edges is likely due to impurities present). Figures \ref{fig:a7}(e)-(h) show DPC reconstructions (in phase units, though remember that source blurring has not been removed) from the corresponding, simultaneously-acquired segmented detector STEM images. On the whole they show images which can be directly interpreted comparably well to the HAADF images. In particular, both readily allow reliable graphene layer counting \cite{cooper2014atomic}.\footnote{This is one instance where the rather large degree of source blurring helps, giving a sense of visual uniformity to the different layers, though if images were obtained with higher resolution one could simply blur in post-processing to aid in counting layers.}
Being more dose efficient, the signal-to-noise ratio in the DPC STEM images is arguably better than that in the HAADF images. However, some residual artefacts are visible, mostly in the form of slowly varying intensity across the full image, such as the darker regions in the corners of \Fig \ref{fig:a7}(g), and the bright region in the upper-right corner (a vacuum region) in \Fig \ref{fig:a7}(h). In what follows we elaborate on how these images have been processed and seek to better understand the source of the artefacts.

Evaluating \Eq (\ref{eq:07}) or (\ref{eq:12}) as written via discrete Fourier transforms implicitly assumes periodic boundary conditions, which is not true of the fields of view in the HAADF images in \Fig \ref{fig:a7}(a)-(d). Ishizuka \etal \cite{ishizuka2017boundary} present one approach to handling this. We take a slightly different approach. The solution of \Eq (\ref{eq:07}) or (\ref{eq:12}) amounts to solving the Poisson equation $\nabla^2 F = \partial_x I_x + \partial_y I_y$. The periodic-boundary-condition-assuming solution of \Eq (\ref{eq:07}) or (\ref{eq:12}) does not satisfy the boundary conditions implied by the experimentally measured values of $\partial_x F = I_x$ and $\partial_y F = I_y$ on the horizontal and vertical boundaries, respectively. However, it does satisfy the Poisson equation at every point in the image interior, and therefore the periodic-boundary-condition-assuming solution of \Eq (\ref{eq:07}) or \Eq (\ref{eq:12}) is only in error by the unique solution of Laplace's equation that satisfies boundary conditions equal to the difference between those measured and those of the periodic-boundary-condition-assuming solution. Solving Laplace's equation in rectangular coordinates via (semi-analytical) series expansion \cite{pinchover2005introduction} provides a fast route for extending the Fourier approach to handle non-periodic boundary conditions. This approach has been used in generating \Figs \ref{fig:a7}(e)-(h). Those figures use the ideal first moment analysis of Eq. (\ref{eq:07}), but we shall subsequently argue that it may not be quite as good an approximation for these large fields of view as it was for the unit-cell average analysis of the previous section.

\begin{figure}[htb!]
\begin{center}
  \includegraphics*[width=1.0\columnwidth]{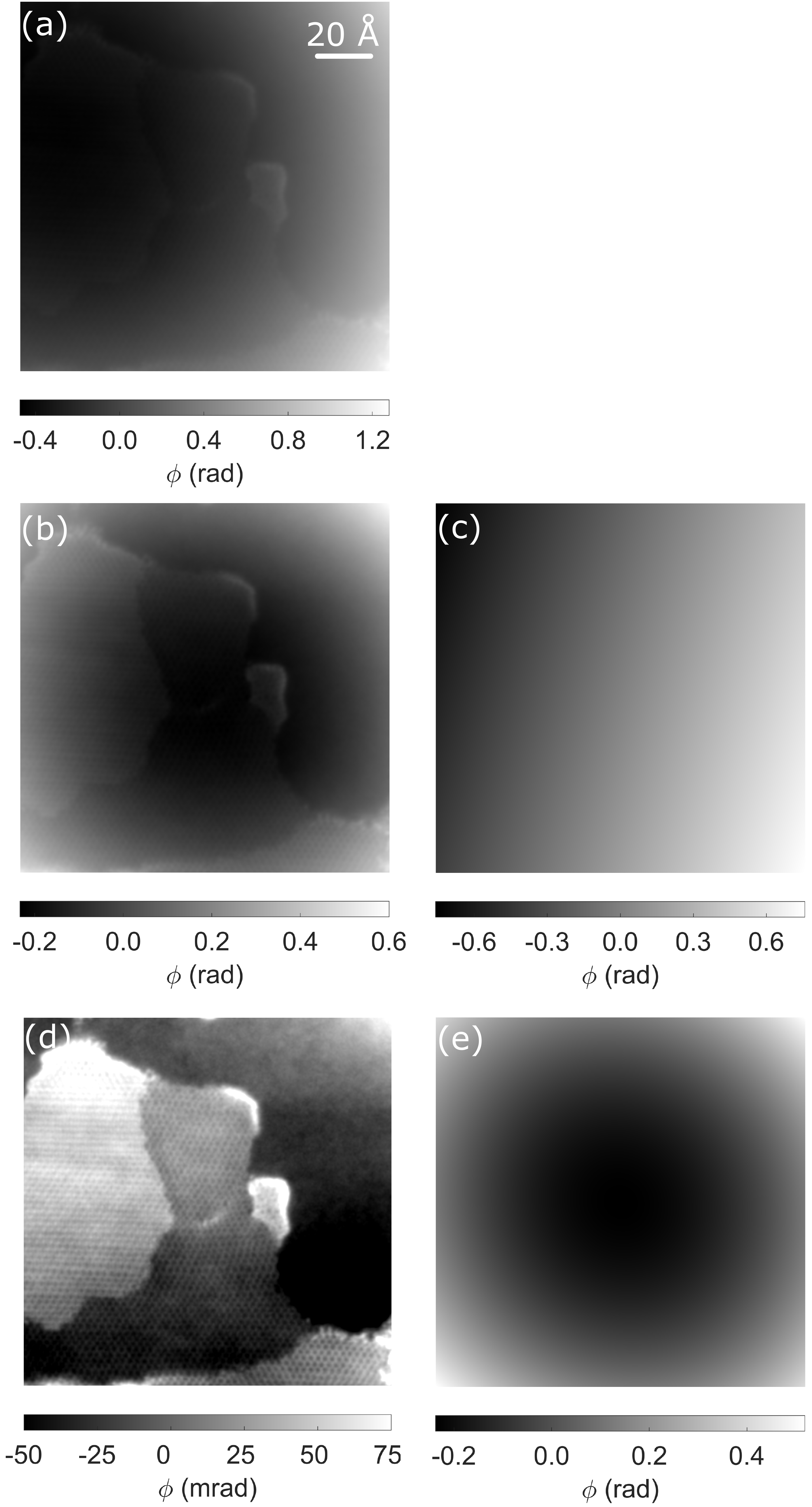}
\caption{(a) Reconstructed phase (source blurring effects included) based on the ideal first moment analysis of Eq. (\ref{eq:07}) modified to handle the non-periodic boundary conditions applied to the non-periodic multilayer graphene sample of \Figs \ref{fig:a7}(d) and (h) using the as-recorded data. (b) Phase obtained by the same reconstruction method after setting the mean of the first moment images to zero. (c) Difference between (a) and (b). (d) Phase obtained by the same reconstruction method after both setting the mean of the first moment images to zero, and further subtracting a best-fit linear function to low-pass-filtered approximate first moment images (with the phase display range truncated slightly to improve the visibility of structure in the mid-range). (e) Difference between (b) and (d).
 \label{fig:a8}}
\end{center}
\end{figure}

If the reconstruction method just described for handling non-periodic boundary conditions is applied directly to the raw segmented detector STEM images for the field of view in \Fig \ref{fig:a7}(d), the result is not that of \Fig \ref{fig:a7}(h) but rather that of \Fig \ref{fig:a8}(a), in which the sample structure is highly obscured by a slowly-varying phase variation across the full image.\footnote{Large errors in low frequency components are a consequence of the poor contrast transfer of low frequency features into the DPC-STEM image and the concomitant potential of the \Eq (\ref{eq:07}) reconstruction method to amplify small errors in these components through the division by near-zero values occasioned by the behaviour of the denominator near the origin in ${\bf K}$-space.} The primary reason for this is that though the variation in the first moment is well-approximated by the segmented detector, it is very difficult to align the centre well enough for the first moment even in vacuum to be truly zero. However, this is easy to remedy by simply subtracting the average of the first moment images from the first moment images prior to phase reconstruction (\ie zeroing their mean). The result of doing that is shown in \Fig \ref{fig:a8}(b). Figure \ref{fig:a8}(c) shows the difference between \Fig \ref{fig:a8}(a) and \Fig \ref{fig:a8}(b), which, as expected from integrating a constant, has the form of a linear ramp. Even though the error in centring was small, its integration across the full image amounts to a significant phase ramp. 

This only goes part way to explaining why the sample structure in \Fig \ref{fig:a8}(a) is highly obscured: the sample structure in \Fig \ref{fig:a8}(b) is still appreciably obscured, now by a slowly varying, concave phase variation. This is attributed to a failure to achieve complete tilt-shift purity, by which we mean that over the large field of view of this scan there is some gradual displacement of the bright field disk (beam tilt) due to the probe scanning alone (\ie independent of the sample).\footnote{This interpretation is reinforced by the observation that the effect becomes steadily less pronounced as the field of view gets smaller --- see \Fig \ref{fig:a7}. Also, similar non-isoplanatic sources of error can affect phase contrast imaging in TEM due to helical electron trajectories, which can be compensated using STEM controls \cite{eades2006obtaining}.} On a pixel detector, this sort of artefact can be identified and corrected in post-processing by recording a scanning diffraction image in the absence of any sample.\footnote{Empty scans are also necessary for other phase contrast imaging techniques, such as removing phase distortions arising from bi-prism charging in off-axis holography \cite{lichte2007electron}.} Given how weakly scattering graphene is, we can do something analogous for the segmented detector data as follows. The approximate first moment images are low-pass filtered (using a Gaussian filter with standard deviation width 1.0 {\AA}) to suppress atomic-scale contrast while having minimal effect on the longer range contrast. Then, a linear function is fit to both first moment images. Subtracting this linear fit from the original approximate first moment images (\ie without low-pass filtering) and reconstructing the phase produces the result in \Fig \ref{fig:a8}(d), where now the sample structure is much more clearly evident. Figure \ref{fig:a8}(e) shows the difference between \Fig \ref{fig:a8}(b) and \Fig \ref{fig:a8}(d), which, as expected from integrating a linear function, has the form of a parabolic surface. 

The low-pass filtering does not eliminate longer range sample contributions from the image that may impact on the fitting. The potential consequences of this are evident in the phase reconstruction in \Fig \ref{fig:a7}(g), where, despite applying the correction steps of \Fig \ref{fig:a8}, the reconstruction contains dark contrast in all image corners. This occurs because our procedure to correct for the imperfect tilt-shift purity ends up over-fitting the slowly-varying, concave phase variation because the genuine structure---a vacuum hole surrounded by a region of monolayer graphene itself surrounded by a region of bilayer graphene---has some genuine longer range concavity. This problem could be ameliorated by excising the layer edges from the fitting region, or an iterative correction procedure guided by some understanding of the structure in the image.

There is also no clear \emph{a priori} reason for believing the lack of tilt-shift purity to be perfectly linear, though being both smooth and a subtle effect means that is a good first approximation. There is some residual structure, particularly in the vacuum region in the upper right of \Fig \ref{fig:a8}(d), that suggests some artefacts remain. However, they no longer obscure the structure of interest and we shall presently show that they may have other origins.

The repeat-unit averaging carried out before the analysis in section \ref{sec:repunitavg} not only guaranteed periodic boundary conditions, it also helped improve the signal-to-noise ratio for the weakly scattering graphene sample considerably by averaging out much of the shot noise and scan distortion. However, close inspection of the HAADF images in \Fig \ref{fig:a7}(a)-(d) suggest both effects are appreciable in these large field of view datasets. Indeed, though DPC is dose efficient \cite{lazic2016phase,yucelen2018phase}, for a weakly scattering sample like graphene it is important to appreciate that while the count rate in the HAADF image is low there is also no background on that signal, whereas in segmented detector STEM images from the bright field region the count rate may be high but consists of a small signal with a large background \cite{hovden2012efficient}. To attempt to separate the impact of different contributing factors, we turn again to simulation.

\begin{figure*}[p!]
\begin{center}
  \includegraphics*[width=1.8\columnwidth]{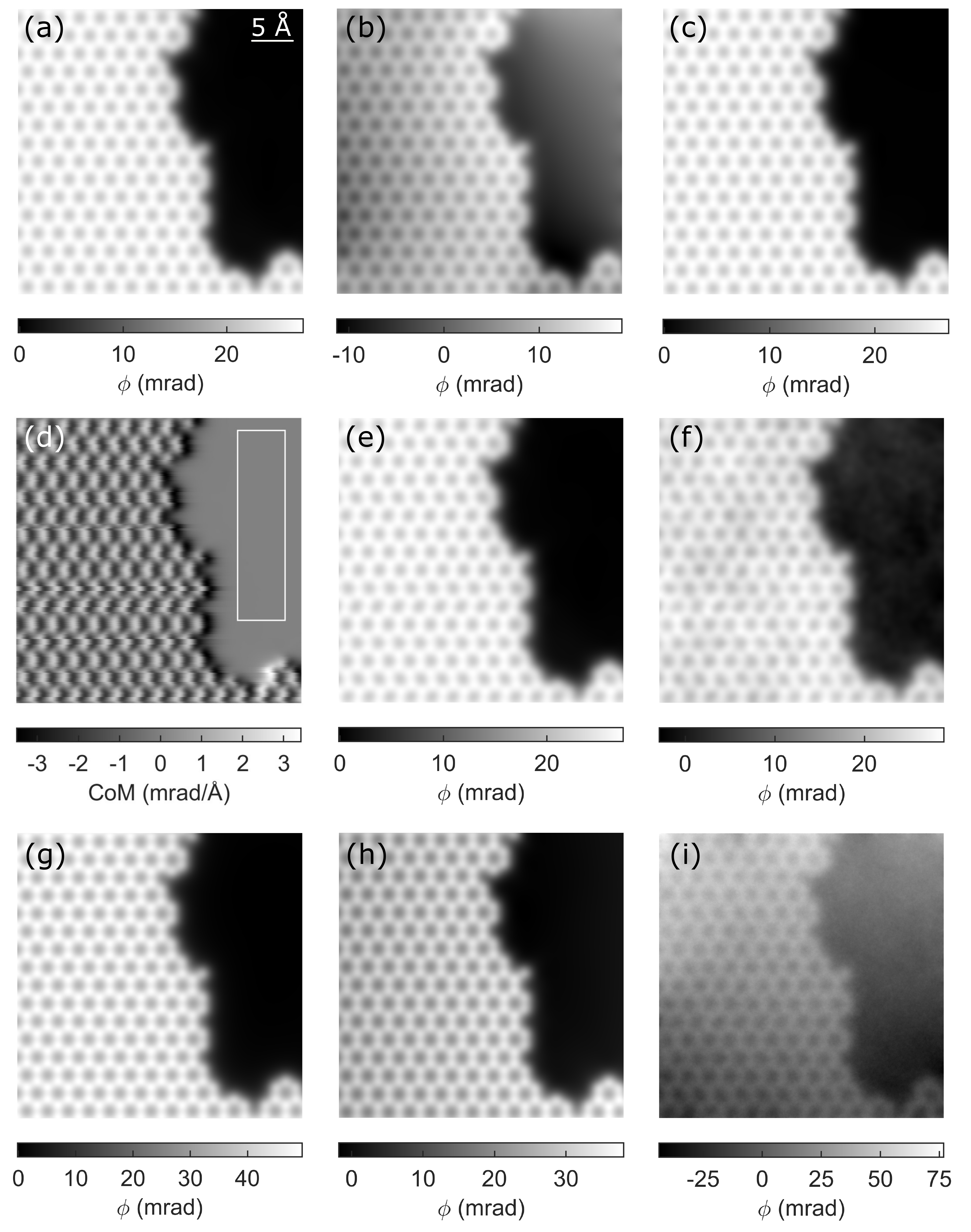}
\caption{(a) Simulated potential of monolayer graphene with a hole in the upper right making the image non-periodic, convolved with the source distribution as per the left side of \Eq (\ref{eq:12}). (b) Reconstruction from simulated DPC STEM images assuming periodic boundary conditions via \Eq (\ref{eq:12}) evaluated using periodic boundary conditions. (c) Reconstruction from simulated DPC STEM images which handles the non-periodic boundary conditions. (d) Centre-of-mass image in the $x$-direction, including a scan distortion shearing effect. The white rectangle indicates the vacuum region used subsequently to compare the noise level. Reconstructions from simulated DPC STEM images which handle the non-periodic boundary conditions for (e) scan distortions but no noise and (f) shot noise corresponding to a dose of $1.6 \times 10^6 \; {\rm e}^-/{\rm {\AA}}^2$. (g), (h) and (i) plot equivalent quantities to (a), (c) and (f) for the ideal first moment analysis of \Eq (\ref{eq:07}).
 \label{fig:a9}}
\end{center}
\end{figure*}

Figure \ref{fig:a9}(a) shows the projected potential of a model structure reminiscent of that of the experimental data in \Figs \ref{fig:a7}(b) and (f). Because we wish to compare it to simulated DPC reconstructions, \Fig \ref{fig:a9}(a) includes the convolution with blurring terms as per the left side of \Eq (\ref{eq:12}). This produces a low contrast within the graphene region, essentially equal to that of the potential data in \Figs \ref{fig:a5}(d) and (e). However, the presence of the hole in the upper-right portion of the image makes for a larger phase range overall. Mean phase not being extractable from DPC reconstructions, for comparison purposes we apply an additive offset such that the mean phase is set to zero within a rectangle in the vacuum region (location shown in \Fig \ref{fig:a9}(d)).

Figure \ref{fig:a9}(b) shows the phase reconstructed from simulated segmented detector STEM images assuming perfect data (\ie no noise or scan distortions) using the approximate first moment, weak phase object approximation of \Eq (\ref{eq:12}), and performing the reconstruction assuming periodic boundary conditions, \ie evaluating \Eq (\ref{eq:12}) via discrete Fourier transform. This is seen to introduce a spurious, slowly-varying phase variation across the reconstruction, even though the considerations about detector centring and tilt-shift purity discussed in reference to \Fig \ref{fig:a8} do not apply. By comparison, \Fig \ref{fig:a9}(c) performs a reconstruction that handles the nonperiodic boundary conditions, resulting in no perceptible slowly-varying phase component. In particular, the phase reconstructed in the hole is very flat.

Figure \ref{fig:a9}(d) shows a simulated $I_{{\rm DPC},x}$ image to which scan noise has been added as a line-by-line shearing achieved by horizontally shifting successive horizontal (fast scan direction) lines by a random number of pixels. The size of the shifts was determined by three parameters – one controlling a small contribution to the shift that varies from line to line, one controlling a larger shift that varies only after multiple lines, and one controlling a variation in that multiple number of lines for which the larger shift remains fixed – all chosen by trial-and-error to give a visual degree of scan distortion reminiscent of that visible in the HAADF image in \Fig \ref{fig:a7}(b). Figure \ref{fig:a9}(e) shows the phase reconstructed from simulated DPC-STEM images that include this scan distortion. The lower frequency components of the scan noise are visible as slight structural distortion in the reconstructed phase, but the phase range is only minimally altered from that in \Fig \ref{fig:a9}(c). This can be largely explained by the high degree of source blurring of the present experiment, such that the lowest-order Fourier coefficients dominate over the higher-order Fourier coefficients (see \Fig \ref{fig:a5}(c)). O'Leary \etal have recently shown that scan distortion correction applied to multiple scans from the same region improves the precision of DPC measurement \cite{o2021increasing}.

Figure \ref{fig:a9}(f) shows the phase reconstructed from simulated 
segmented detector STEM images that (in addition to scan noise) include shot noise corresponding to a total dose of $1.6 \times 10^6 \; {\rm e}^-/{\rm {\AA}}^2$ or 2300 electrons per probe position given the step size used here. This value was obtained by trial and error such that standard deviations of the noise fluctuations within the vacuum region indicated by the white rectangle in \Fig \ref{fig:a9}(d) is comparable to that in a similar vacuum region in the experimental images.\footnote{Note that because the DPC-STEM image is a processed image, its noise distribution is a consequence of propagating the shot noise in each detector segment image through the centre of mass evaluation of using \Eq (\ref{eq:08}) in \Eq (\ref{eq:03}). But because the number of pixels in the comparison region is large, the distribution of noise values is essentially Gaussian, and because the centre of mass is inherently on a quantitative scale (even when the number of electron counts are unknown), the trial-and-error comparison of what dose makes the experimental and simulated distributions agree is straightforward.} This is high even by materials science standards, but necessary because graphene is a weak scatterer.\footnote{Yang \etal \cite{Yang_2014_PN} show that pixel detectors can achieve the same contrast for somewhat lower dose.} Figure \ref{fig:a9}(f) shows that this introduces ``clouding'', a diffuse signal component with texture across multiple length scales. This is a common feature of Fourier-based phase retrieval \cite{paganin2004quantitative,clark2019high}, arising from the low frequency noise enhancement that results here from the Fourier approach to the inverse Laplacian operator. Nevertheless, it has had only a modest impact on the overall phase range.

Figures \ref{fig:a7} and \ref{fig:a8} were processed via the ideal first moment analysis of \Eq (\ref{eq:07}): despite being only an approximation for data from a segmented detector, section \ref{sec:repunitavg} suggested it to be a good approximation. Previous work suggests it can be a smaller source of error than that of assuming the phase object approximation applies to thicker samples \cite{close2015towards}. However, we now show that it can introduce some undesirable features for large field of view imaging, which the weak phase object approximation analysis of \Eq (\ref{eq:12}) avoids. Figures \ref{fig:a9}(g), (h) and (i) are the ideal first moment analysis (\ie \Eq (\ref{eq:07})) equivalents of \Figs \ref{fig:a9}(a), (c) and (f), that is to say the (source blurred) potential, the reconstruction from noise-free data (handling the non-periodic boundary conditions), and the reconstruction in the presence of both scan and shot noise. The phase range of \Fig \ref{fig:a9}(g) is larger than \Fig \ref{fig:a9}(a) because it does not involve the additional convolution of \Eq (\ref{eq:11}) needed in the weak phase object approximation to yield the $x$ and $y$-directed gradients of the same function. However, the noise free reconstruction in \Fig \ref{fig:a9}(h) has a reduced range relative to that of the reference potential in \Fig \ref{fig:a9}(g). This comes not from the monolayer graphene region, which consistent with section \ref{sec:repunitavg} is well reconstructed despite the segmented detector data yielding only an approximate first moment. Rather, the discrepancy comes through a reduction in the difference between the graphene and vacuum phases in the reconstruction, resulting from the segmented detector underestimating the first moment at the graphene edge. This can be understood from the contrast transfer function ratio in \Fig \ref{fig:a4}(h): while the ratio is approximately constant at the first order frequencies in pure graphene, it differs at other frequencies, such as those needed to describe the sample edge. Furthermore, \Fig \ref{fig:a9}(i) shows that the ideal first moment reconstruction is more sensitive to noise. This occurs because the additional convolution of \Eq (\ref{eq:11}) needed in the weak phase object approximation helps to further suppress noise at high frequencies, and, more than affecting the extremes of the phase range, the reduction of shot noise around the boundary improves the robustness of applying the non-periodic boundary conditions.

\section{Discussion and conclusions}

Many experimental factors must be characterised, controlled or otherwise accounted for when seeking to make quantitative comparison between segmented detector DPC STEM experiments and simulation. The present case study used a 16-segment detector and monolayer and few-layer graphene samples to explore the relative impacts of such factors. Coherent and incoherent lens aberrations were seen to be the dominant factors impacting both qualitative agreement, as evident through deviations from expected symmetry, and quantitative agreement, as evident in phase range of the reconstructions. By comparison, on repeat-unit-averaged data the effects of probe-forming aperture, detector camera length, detector orientation, detector mis-centering and modest detector non-uniformity were very small. Indeed, the almost negligible impact of these limitations is surprising, and may be somewhat specific to the present structure: that the contrast transfer function is largely insensitive to modest variations in these quantities at the low order frequencies in graphene does not guarantee the same is true at all frequencies. Nevertheless, carrying out analogous simulation explorations for other structures of interest is straightforward, and the much larger impact of coherent and incoherent lens aberrations suggests that improving their characterisation is the priority for improving the accuracy of quantitative DPC STEM in general. See Ishikawa \etal \cite{ishikawa2021automated} for a recent approach to improving coherent lens aberration correction.

For larger fields of view with non-periodic boundary conditions, our explorations show that care is needed to correct for any detector mis-centring or break down of tilt-shift purity as even small deviations accumulate rapidly over wide fields of view. Relative to that, the effects of noise and scan distortions are less significant.

Much of the findings here would also apply to DPC STEM using pixel detectors, when available and at the expense of much larger data sets, though that detector geometry may facilitate instrument characterisation (\eg readily visualising imperfect tilt-shift purity in a scan in the absence of a sample), or allow more sophisticated phase reconstruction strategies \cite{jiang2018electron,chen2021electron}. Nevertheless, segmented detectors also appear to be proliferating, with advantages including rapid acquisition with simultaneous visualisation of outputs. Explorations like that presented here can help clarify where further effort in instrument characterisation will yield the most benefit. They can also help guide the design of future such segmented detector DPC experiments, and the development of analysis strategies to maximise the accuracy and precision of segmented detector DPC STEM imaging.

\section*{Acknowledgements}

This research was supported under the Australian Research Council's Discovery Projects funding scheme (Project DP160102338).  D.J.T. acknowledges support through an Australian Government Research Training Program Scholarship. This project has received funding from the European Union's Horizon 2020 research and innovation programme under the Marie Sklodowska-Curie grant agreement No 891504. R.I., T.S., and N.S. acknowledge the support from JSPS KAKENHI Grant No. 19H05788. R.I. acknowledges the support from JSPS KAKENHI Grant No. 21H01614 and JST PRESTO. T.S. and N.S. acknowledge support from the JST SENTAN Grant No. JPMJSN14A1 and JSPS KAKENHI Grant No. 20H05659.

\end{document}